\documentstyle[12pt,psfig]{article}
\catcode`\@=11
\long\def\@makefntext#1{
\protect\noindent \hbox to 3.2pt {\hskip-.9pt

$^{{\ninerm\@thefnmark}}$\hfil}#1\hfill}		%CAN BE USED

 \def\@makefnmark{\hbox to 0pt{$^{\@thefnmark}$\hss}}  %ORIGINAL

\def\ps@myheadings{\let\@mkboth\@gobbletwo
\def\@oddhead{\hbox{}
\rightmark\hfil\ninerm\thepage}

\def\@oddfoot{}\def\@evenhead{\ninerm\thepage\hfil
\leftmark\hbox{}}\def\@evenfoot{}
\def\sectionmark##1{}\def\subsectionmark##1{}}

\newcounter{sectionc}\newcounter{subsectionc}\newcounter{subsubsectionc}
\renewcommand{\section}[1] {\vspace{0.6cm}\addtocounter{sectionc}{1}

\setcounter{subsectionc}{0}\setcounter{subsubsectionc}{0}\noindent
	{\bf\thesectionc. #1}\par\vspace{0.4cm}}
\renewcommand{\subsection}[1]
{\vspace{0.6cm}\addtocounter{subsectionc}{1}

	\setcounter{subsubsectionc}{0}\noindent

	{\it\thesectionc.\thesubsectionc. #1}\par\vspace{0.4cm}}
\renewcommand{\subsubsection}[1]
{\vspace{0.6cm}\addtocounter{subsubsectionc}{1}
	\noindent
{\rm\thesectionc.\thesubsectionc.\thesubsubsectionc.
	#1}\par\vspace{0.4cm}}

\newcounter{appendixc}
\newcounter{subappendixc}[appendixc]
\newcounter{subsubappendixc}[subappendixc]

\renewcommand{\appendix}[1] {\vspace{0.6cm}
        \refstepcounter{appendixc}
        \setcounter{figure}{0}
        \setcounter{table}{0}
        \setcounter{equation}{0}
        \renewcommand{\thefigure}{\Alph{appendixc}.\arabic{figure}}
        \renewcommand{\thetable}{\Alph{appendixc}.\arabic{table}}
        \renewcommand{\theappendixc}{\Alph{appendixc}}

\renewcommand{\theequation}{\Alph{appendixc}.\arabic{equation}}
        \noindent{\bf Appendix \theappendixc #1}\par\vspace{0.4cm}}

\def\abstracts#1{{
	\centering{\begin{minipage}{30pc}\tenrm\baselineskip=12pt\noindent
	\centerline{\tenrm ABSTRACT}\vspace{0.3cm}
	\parindent=0pt #1
	\end{minipage}}\par}}

\renewenvironment{thebibliography}[1]
	{\begin{list}{\arabic{enumi}.}
	{\usecounter{enumi}\setlength{\parsep}{0pt}
\setlength{\leftmargin 1.25cm}{\rightmargin 0pt}
	 \setlength{\itemsep}{0pt} \settowidth
	{\labelwidth}{#1.}\sloppy}}{\end{list}}

\newcounter{itemlistc}
\newcounter{romanlistc}
\newcounter{alphlistc}
\newcounter{arabiclistc}

\newcommand{\fcaption}[1]{
        \refstepcounter{figure}
        \setbox\@tempboxa = \hbox{\tenrm Fig.~\thefigure. #1}
        \ifdim \wd\@tempboxa > 6in
           {\begin{center}
        \parbox{6in}{\tenrm\baselineskip=12pt Fig.~\thefigure. #1}
            \end{center}}
        \else
             {\begin{center}
             {\tenrm Fig.~\thefigure. #1}
              \end{center}}
        \fi}

\newcommand{\tcaption}[1]{
        \refstepcounter{table}
        \setbox\@tempboxa = \hbox{\tenrm Table~\thetable. #1}
        \ifdim \wd\@tempboxa > 6in
           {\begin{center}
        \parbox{6in}{\tenrm\baselineskip=12pt Table~\thetable. #1}
            \end{center}}
        \else
             {\begin{center}
             {\tenrm Table~\thetable. #1}
              \end{center}}
        \fi}

\def\@citex[#1]#2{\if@filesw\immediate\write\@auxout
	{\string\citation{#2}}\fi
\def\@citea{}\@cite{\@for\@citeb:=#2\do
	{\@citea\def\@citea{,}\@ifundefined
	{b@\@citeb}{{\bf ?}\@warning
	{Citation `\@citeb' on page \thepage \space undefined}}
	{\csname b@\@citeb\endcsname}}}{#1}}

\newif\if@cghi
\def\cite{\@cghitrue\@ifnextchar [{\@tempswatrue
	\@citex}{\@tempswafalse\@citex[]}}
\def\citelow{\@cghifalse\@ifnextchar [{\@tempswatrue
	\@citex}{\@tempswafalse\@citex[]}}
\def\@cite#1#2{{$\null^{#1}$\if@tempswa\typeout
	{IJCGA warning: optional citation argument
	ignored: `#2'} \fi}}

\def\fnt#1#2{\footnotetext{\kern-.3em
	{$^{\mbox{\sevenrm #1}}$}{#2}}}

 1
 1
\font\twelveit=cmti10 scaled\magstep 1

\font\tenbf=cmbx10
\font\tenrm=cmr10
\font\tenit=cmti10

\font\ninerm=cmr9

\begin{document}

%%%%%%%%%%%%%%%%%%%%%%%%%%Our Own Definitions%%%%%%%%%%%
%========================================================================
\newcommand{\thspace}{\kern.08333em}
\newcommand{\sst}{\scriptstyle}
\newcommand{\bd}{B_d^0}
\newcommand{\bdb}{\overline{B_d^0}}
\def\beq{\begin{equation}}
\def\eeq{\end{equation}}
\def \beq{\begin{equation}}
\def \eeq{\end{equation}}
\def \beqn{\begin{eqnarray}}
\def\eeqn{\end{eqnarray}}
\def\Bbar{\overline{B}}
\def\bbar{\overline{B^0}}
\def\Kbar{\overline{K}}
\def\sss{\scriptscriptstyle}
\def\barp{{\raise.35ex\hbox{${\sss (}$}}---{\raise.35ex\hbox{${\sss )}$}}}
\def\bdbarp{\hbox{$B_d$\kern-1.4em\raise1.4ex\hbox{\barp}}}
\def\bsbarp{\hbox{$B_s$\kern-1.4em\raise1.4ex\hbox{\barp}}}
\def\dcp{D^0_{\sss CP}}
\def\s{\sqrt{2}}
\def\st{\sqrt{3}}
\def\sx{\sqrt{6}}
\def\v#1#2{V_{#1#2}}
\def\vc#1#2{V^*_{#1#2}}
\def\ruf#1{\raise.3ex\hbox{$#1$\kern-.75em\lower1ex\hbox{$\sim$}}}
\def\lsim{\ruf<}
%
%=========================================================================

\rightline{UdeM-GPP-TH-95-20}
\rightline{hep-ph/9502355}
\rightline{February 1995}
\bigskip
\centerline{\tenbf HOW TO USE SU(3)-FLAVOR SYMMETRY TO EXTRACT }
\baselineskip=22pt
\centerline{\tenbf CP VIOLATING PHASES AND STRONG PHASE SHIFTS
AT CLEO\footnote{Based on a talk presented at the Mexican School of Particles
and Fields in Villahermosa, Tabasco, October 1994, and to appear in the
proceedings published by World Scientific.}}
%\baselineskip=16pt
%\centerline{\tenbf }
%\centerline{\ninerm (For 20\% Reduction to 8.5 $\times$ 6 in Trim Size)}
\vspace{0.8cm}
\centerline{\tenrm OSCAR F. HERN\'ANDEZ}
\baselineskip=13pt
\centerline{\tenit e-mail: oscarh@lps.umontreal.ca}
\centerline{\tenit Laboratoire de Physique Nucl\'eaire, Universit\'e de
Montr\'eal}
\baselineskip=12pt
\centerline{\tenit Montr\'eal, QC, H3C 3J7, Canada}
\vspace{0.9cm}
\abstracts{
After a short introduction to the SM picture of CP violation, we
discuss recent work on how to use SU(3) flavor symmetry, along with
some dynamical approximations, to extract the CKM weak phases and the
strong rescattering phases from experimental measurements alone.  This
suprising wealth of information depends on our two strongest
assumptions: SU(3) invariance, and the relative unimportance of
exhange and annihilation diagrams.  We discuss soon to be measured
decay rate measurements that will test the validity of these
assumptions.  }

\vfil
%\vspace{0.8cm}
\rm\baselineskip=14pt
\section{Introduction}
\vspace{-0.7cm}
\subsection{Charge and Parity Conjugation}
\vspace{-0.35cm}

If asked the reason for the apparent left-right asymmetry in our
bodies  (location of our heart, liver, the asymmetry in our face, etc.)
most of us would probably say that it has to do with random initial
conditions, either evolutionary or developmental.  Thus despite
the fact that in our daily lifes things are not completely
left-right symmetric, before 1957 physicists took it for granted that
the fundamental laws of physics were parity invariant. It came as
quite a shock to everyone when in 1957 a left-right asymmetry was
discoverd in beta, pion, and muon decays\cite{noparity}.  Since
these same experiments established an asymmetry between the
decays of postive and negative particles, an absolute (versus relative)
difference between positive and negative charges was established.

Well then, if we can't have P-invariance or C-invariance how about
CP-invariance. CP, along with P and C conjugation is a discrete
symmmetry of the Poincar\'e group, which was respected by the P
and C violating beta, pion and muon decay experiments, and by all
the macroscopic laws of physics.\footnote{\ninerm Of course on the
macroscopic level P and C are also good symmetries. So billiards in a
mirror is indistinguishable from billiards in real life. And
billiards with positively charged balls is in indistiguishable from
billiards with negatively charged balls and they are both also
equivalent to billiards with oppositely charged balls in the mirror.}
But then in 1964 came the second surprise\cite{nocp}. The CP odd
state $K_L$ decayed once every couple thousand times into the CP
even state of two pions.

\begin{figure}
\centerline{\psfig{figure=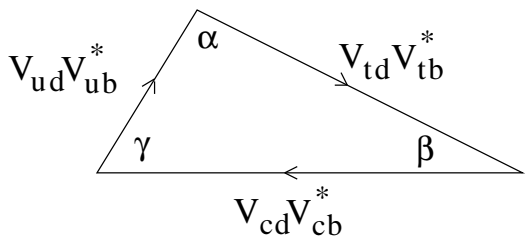,height=3.0cm,angle=90}}
\caption{\tenrm\baselineskip=12pt The unitarity triangle. }
\label{unitaritytriangle}
\end{figure}

%\vspace{0.4cm}
%\vspace{0.3cm}
\leftline{\twelveit 1.2. The Standard Model Unitary Triangle}
%\vspace{0.4cm
\vspace{1pt}
So far the Kaon system is the only system where CP violation has
been observed.  In the future we will have a statistically large enough
sample of $B$-mesons that their rare decays will also become a
crucial testing ground for our ideas about CP violation.  The
Standard Model (SM) picture of CP violation, is based on phases in
the Cabibbo-Kobayashi-Maskawa (CKM) matrix\cite{CKM}. In
studying CP violation in the $B$ system, it is convenient to use an
approximate form of the CKM matrix, due to Wolfenstein
\cite{Wolfenstein}. This approximation is based on the observation
that the elements of the CKM matrix obey a hierarchy in
powers of the Cabibbo angle, $\lambda=0.22$:
\beq
\left(\begin{array}{ccc}
V_{ud} & V_{us} & V_{ub} \\
V_{cd} & V_{cs} & V_{cb} \\
V_{td} & V_{ts} & V_{tb}\end{array}\right)
\sim
\left(\begin{array}{ccc}
1-{1\over 2}\lambda^2 & \lambda & |V_{ub}|\exp (-i\gamma ) \\
-\lambda  & 1- {1\over 2}\lambda^2 & A \lambda^2 \\
|V_{td}|\exp (-i\beta ) & -A \lambda^2 & 1
\end{array}\right).
\eeq
Here, $A$ is a parameter of $O(1)$, and $|V_{ub}|$ and $|V_{td}|$
are terms of order $\lambda^3$. In this approximation, the only non-
negligible complex phases appear in the terms $V_{ub}$ and
$V_{td}$. Unitarity of the CKM matrix implies, among other things,
the orthogonality of the first and third columns:
\beq
\v u d \vc u b + \v c d \vc c b + \v t d \vc t b = 0 ~.
\eeq
This relation can be represented as a triangle in the complex plane
(the unitarity triangle), as shown in Fig.~\ref{unitaritytriangle}. In
the Wolfenstein approximation, the angles in the unitarity triangle
are given by $\beta =-{\rm Arg}(\v td)$, $\gamma ={\rm Arg}(\vc
ub)$, and $\alpha =\pi -\beta -\gamma $\cite{CPangles}. The SM
picture of CP violation can thus be tested by independently
measuring the three angles $\alpha$, $\beta$ and $\gamma$ and
seeing (i) that they are all different from $0$ or $\pi$, and (ii) that
they add up to $\pi$ radians.

\section{Measuring CP Violation in $B$ Decays}
\subsection{Direct CP Violation}
\begin{figure}
\centerline{\psfig{figure=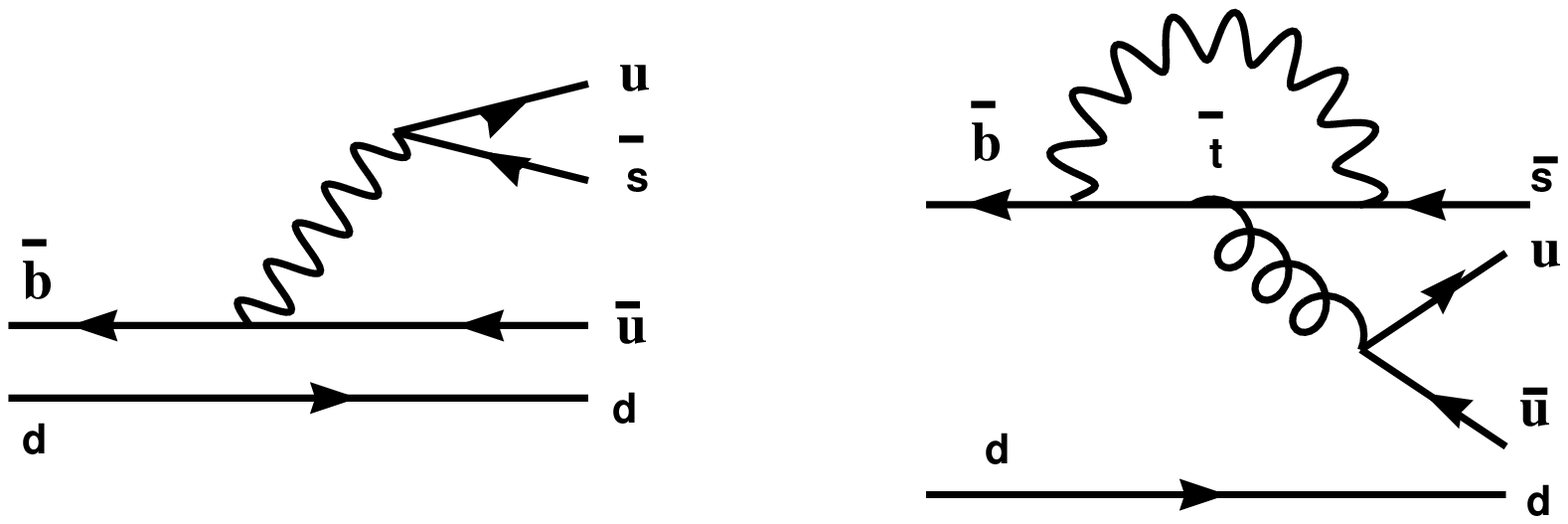,height=4.0cm,angle=0}}
\caption{\tenrm\baselineskip=12pt Diagrams contributing to the
process $\bd\to\pi^-K^+$.}
\label{btopik}
\end{figure}

The most straightforward way to detect CP violation in $B$ decays
would be to observe a difference between the decay of a $B$ meson
to a final state $f$ and the CP-conjugate process:
\beq
\label{direct}
\Gamma(B\to f) \ne \Gamma(\bar B \to \bar f)
\eeq
Let's take for example the case $B^0\to \pi^- K^+$.  In the quark
model representation of the mesons, this process is the sum of a tree
and a penguin diagram in Fig.~\ref{btopik}, dressed with gluons
in all possible places.  It is the contribution of these soft gluons which
we can not calculate.  Still the amplitude for this process is the sum
of two complex numbers, one coming from the tree diagram $T$ and
the other from the penguin diagram $P$.  The phase associated with
each diagram has a ``weakÆÆ contribution and a ``strongÆÆ
contribution.  The weak phases $\phi$ are due to the CKM matrix
elements and they change sign in the CP conjugate process, whereas
the strong phases $\delta$ are due to hadronization and final state
rescattering effects and they do are the same for both the original
decay and the CP conjugate process.  This is because CP violation
does not occur in the strong interactions (as upper bounds on the
neutron electric dipole moment show), but only in the weak sector.

Thus we write
\begin{eqnarray}
A(B^0\to \pi^- K^+) & = &T \, {\rm e}^{i\phi_T} \, {\rm
e}^{i\delta_T}
            + P \, {\rm e}^{i\phi_P} \, {\rm e}^{i\delta_P}~, \nonumber
\\
A(B^0\to \pi^+ K^-) & = & T \, {\rm e}^{-i\phi_T} \, {\rm
e}^{i\delta_T}
		       + P \, {\rm e}^{-i\phi_P} \, {\rm e}^{i\delta_P}~.
\end{eqnarray}
(In this case $\phi_T=arg(V_{ub}^* V_{us})=\gamma$ and
$\phi_P=arg(V_{tb}^* V_{ts})=\pi$.) It is straightforward to show
that the difference in the decay rates is
\beq
\Gamma(B^0\to \pi^- K^+)-\Gamma(B^0\to \pi^+ K^-)
\sim \sin(\phi_T-\phi_P)\sin(\delta_T-\delta_P).
\eeq
Note that, although this rate asymmetry is proportional to
$\sin(\phi_T-\phi_P) \sim \sin\gamma$, it also depends on the strong
phase difference $\sin(\delta_T-\delta_P)$. The problem is that these
strong phases are incalculable. Thus, a measurement of the rate
asymmetry in $B^0\to\pi^- K^+$ does not provide {\it clean}
information on the CKM phases. This is true of all processes which
involve direct CP violation. However, we will soon see how to use
flavor SU(3) to separate the weak and the strong phases, so that
direct CP-violation measurements can in fact be used to extract the
weak phases cleanly.

%\vspace{0.4cm}
\vspace{0.3cm}
\leftline{\twelveit 2.2  Indirect CP Violation}
%\vspace{0.4cm
\vspace{1pt}
Suppose on chooses a final state $f$ to which both $B^0$ and
$\bbar$ can decay. Then due to $B^0-\bbar$ mixing, there will be
interference between the two amplitudes $B^0\to f$ and
$B^0\to\bbar\to f$ which allows us to observe CP violation.

In order to be able to obtain clean CKM phase information, it is a
necessary requirement that only one weak amplitude contribute to the
decay otherwise direct CP violation is introduced, ruining the
cleanliness of the measurement.  This is in fact the case for many
$B$ decays such as $\bdbarp\to\pi^+\pi^-$, as shown in
Fig.~\ref{btopipi}. Here, the tree diagram has the weak phase $\vc u
b \v u d$ ($\sim \gamma$), while that of the penguin diagram is $\vc
t b \v t d$ ($\sim \beta$). In other words, in this decay, in
addition to indirect CP violation, direct CP violation is present due
to the interference of the tree and penguin diagrams. The presence of
direct CP violation spoils the cleanliness of the measurement, hence
the term ``penguin pollution.'' Thus a measurement of the CP
asymmetry in this mode does not give access to a CKM phase ($\alpha$
in this case) but rather we measure a quantity that
depends on the weak and strong phases of the tree and penguin
diagrams, as well as on their relative sizes.

\begin{figure}
\centerline{\psfig{figure=btopipi.ps,height=4.0cm,angle=90}}
\caption{\tenrm\baselineskip=12pt Diagrams contributing to the process
$\bd\to\pi^+\pi^-$.}
\label{btopipi}
\end{figure}

All is not lost, however. Even in the presence of penguin diagrams, it
is still possible to cleanly extract the CKM phase $\alpha$ by using
an isospin analysis \cite{isospin}. The idea is to use isospin to
relate the three amplitudes $A(\bd\to\pi^+\pi^-)$,
$A(\bd\to\pi^0\pi^0)$ and $A(B^+\to\pi^+\pi^0)$, and similarly for
the CP-conjugate processes. For all these decays, the final state has
total isospin $I=0$ or 2. In other words, there are two amplitudes
for these decays: $\Delta I=1/2$ and $\Delta I=3/2$. Since there are
two isospin amplitudes, but three $B$-decay amplitudes, there must be
a triangle relation among the $B$ amplitudes. It is:
\beq
{1\over \sqrt{2}} A^{+-} + A^{00} = A^{+0}~.
\eeq
There is a similar relation among the CP-conjugate processes:
\beq
{1\over \sqrt{2}} {\bar A}^{+-} + {\bar A}^{00} = {\bar A}^{-
0}~.
\eeq
Note that the tree diagram has both $\Delta I=1/2$ and $\Delta I=3/2$
pieces, but the penguin diagram is pure $\Delta I=1/2$. We can
isolate the $\Delta I=3/2$ contribution and remove the ``penguin
pollution'' by using the triangle relations above.  One measures the
rates for the three decay processes and their CP conjugate processes
(six in total) and constructs the triangles as in
Fig.~\ref{isotriangles}. (In this figure, the ${\tilde A}$'s are
related to the ${\bar A}$'s by a rotation.) Thus, up to a discrete
ambiguity (since one or both triangles may be flipped upside-down),
this determines $\theta_{+-}$, the penguin pollution. With this
knowledge the angle $\alpha$ can be extracted by measuring CP
violation in $\bd(t)\to\pi^+\pi^-$. Therefore, even in the presence
of penguins, $\alpha$ can be obtained cleanly by using the above
isospin analysis.

\begin{figure}
\centerline{\psfig{figure=isotriangles.ps,height=3.5cm,angle=90}}
\caption{\tenrm\baselineskip=12pt Isospin triangles in $B\to\pi\pi$.}
\label{isotriangles}
\end{figure}

\section{SU(3) Relations Among Amplitudes}  In general, our inability
to calculate strong interaction effects hampers out ability to
cleanly extract a CKM phase from a our measurment of decay
processes.  We have seen that SU(2) isospin allowed us to get around
these problems in mixing-induced CP violation measurements, however
difficult tagging and time dependent measurements are required.
There is a way to use $B\to D K$ decays to obtain clean CP violation
information without tagging. However the triangle that needs to be
constructed is expected to be very thin; two of the sides will be an
order of magnitude longer than the third.  On top of this, only the
angle $\gamma$ can  be extracted this way.

The successful application of isospin symmetry in the $B\to \pi\pi$
analysis begs the question, ``what information does flavor  SU(3)
allow us to extract''. The answer is ``just about everything''
\cite{Zeppenfeld}- \cite{HLGR}. We  will see that, together with a
few simple approximations, SU(3)  symmetry  allows us to obtain all
of the CKM weak phases and all of the strong  phase  shifts from {\it
time-independent} measurements alone without tagging.  Since tagging
and time-dependent measurements are not necessary, our analysis
allows CLEO to scoop the $B$ factory in the search for CP violation
in the $B$ system.

\begin{figure}
\centerline{\psfig{figure=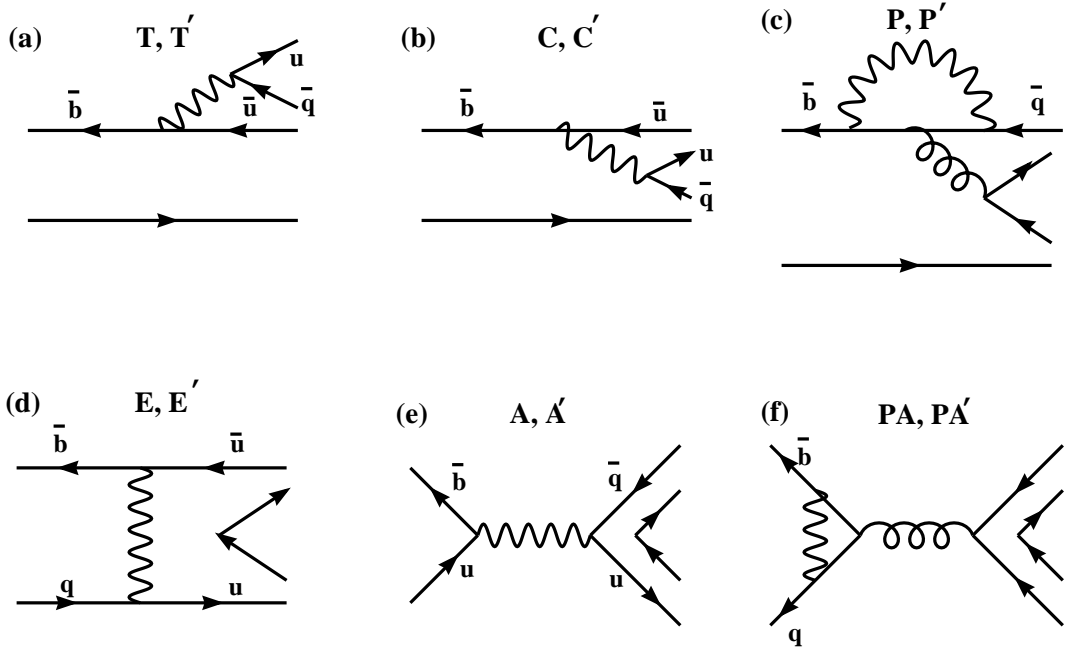,height=9.0cm,bbllx=-5cm}}
\caption{\tenrm\baselineskip=12pt Diagrams describing decays of  $B$
mesons to pairs of light pseudoscalar mesons. Here $\bar q = \bar d$
for  unprimed amplitudes and $\bar s$ for primed amplitudes.
(a) ``Tree'' (color-favored) amplitude $T$ or $T'$;
(b) ``Color-suppressed'' amplitude $C$ or $C'$;
(c) ``Penguin'' amplitude $P$ or $P'$ (we do not show intermediate
quarks and gluons);
(d) ``Exchange'' amplitude $E$ or $E'$; (e) ``Annihilation''
amplitude $A$ or $A'$; (f) ``Penguin annihilation'' amplitude $PA$
or
$PA'$.}
\label{diagrams}
\end{figure}

In going from SU(2) to SU(3) the number of Goldstone bosons
increases from  3 (the $\pi$'s) to 8 with the addition of $K,\Kbar,
K^+, K^-$ and the  $\eta$ (we ignore the $\eta$ from now on because
of its limited  experimental usefulness). Following the conventions
in  Refs.~\cite{Zeppenfeld,GHLR}, we take the $u$, $d$, and $s$
quark  to  transform as a triplet of flavor SU(3), and the $-\bar u$,
$\bar d$,  and  $\bar s$ to transform as an antitriplet. Thus the
$\pi$-mesons and  kaons  form part of an octet and are defined as
$\pi^+ \equiv u \bar d$,  $\pi^0  \equiv (d \bar d - u \bar u)/\s$,
$\pi^- \equiv - d \bar u$, $K^+ \equiv  u  \bar s$, $K^0 \equiv d
\bar s$, $\bar K^0 \equiv s \bar d$ and $K^-  \equiv  - s \bar u$.
The $B$ mesons, which are in the triplet or anti-triplet
representation, are taken to be $B^+ \equiv \bar b u$, $B^0 \equiv
\bar b  d$, $B_s \equiv \bar b s$, $B^- \equiv - b\bar u$, $\Bbar^0
\equiv b  \bar  d$ and $\Bbar_s \equiv b\bar s$.

Consider all the decays of $B$ mesons to pairs of light pseudoscalar
mesons  $\pi\pi$, $\pi K$ and $K {\bar K}$. The amplitudes for these
decays  can be  expressed in terms of the following diagrams (see
Fig.~\ref{diagrams}): a  ``tree'' amplitude $T$ or $T'$, a
``color-suppressed'' amplitude $C$  or  $C'$, a ``penguin'' amplitude
$P$ or $P'$, an ``exchange'' amplitude  $E$ or  $E'$, an
``annihilation'' amplitude $A$ or $A'$, and a ``penguin
annihilation'' amplitude $PA$ or $PA'$. Here an unprimed amplitude
stands  for a strangeness-preserving decay, while a primed
contribution  stands for  a strangeness-changing decay. As noted in
Refs.~\cite{Zeppenfeld,GHLR},  this set of amplitudes is
over-complete. The physical processes of  interest  involve only five
distinct linear combinations of these six terms.

Now comes one of the main points. The diagrams denoted by $E$,  $A$
and $PA$  can be ignored relative to the other diagrams. The reasons
are as  follows.  First, the diagrams $E$ and $A$ are helicity
suppressed by  $(m_{u,d,s}/m_B)$ since the $B$ mesons are
pseudoscalars.  Second,  annihilation and exchange processes, such as
those represented by  $E$, $A$,  $PA$, are directly proportional to a
factor of the $B$-meson wave  function  at the origin. Thus these
diagrams are suppressed by a factor of  $(f_B/m_B)\ruf{<}0.05$
relative to diagrams $T$, $C$ and  $P$ (and  similarly for their
primed counterparts). This suppression should  remain  valid unless
hadronization and rescattering effects are important.  Such
rescatterings could be responsible for certain decays of charmed
particles,  but should be less important for the higher-energy $B$
decays.

Neglecting the contributions of the above diagrams, we are left with
the 6  diagrams $T$, $T'$, $C$, $C'$, $P$ and $P'$. These six
complex  parameters  determine the 13 allowed $B$ decays to states
with pions and kaons,  as  listed in Table 1. This table is derived
by expressing the $B$ into  pseudoscalar decay as graphs in terms of
their quark level  contributions,  keeping track of minus signs and
$\sqrt{2}$ factors in going from  quarks to  mesons. The primed and
unprimed diagrams are not independent, but  are  related by CKM
matrix elements. In particular, $T'/T = C'/C = r_u$,  where  $r_u
\equiv \v u s/ \v u d\approx 0.23$. Assuming that the penguin
amplitudes are dominated by the top quark loop, one has $P'/P =
r_t$, with  $r_t \equiv \v t s/ \v t d$. We therefore have 13 decays
described by  3  independent graphs, implying that there are 10
relations among the  amplitudes. These can be expressed in terms of 6
amplitude  equalities, 3  triangle relations, and one quadrangle
relation.

\begin{table}
\caption{ \tenrm\baselineskip=12pt   The 13 decay
amplitudes in terms of the 8 graphical combinations.  The $\sst
\protect\sqrt{2}(B^+\to\pi^+\pi^0)$ in the $\sst -(T+C)$ column
means that $\sst A(B^+\to\pi^+\pi^0) = -(T+C)/\protect\sqrt{2}$, and
similarly  for  other entries. Processes in the same column can be
related by an  amplitude  equality, e.g.~the amplitudes for $\sst
B^+\to K^+\Kbar^0$ and $\sst  B^0\to  K^0\Kbar^0$ are equal.}
\begin{center}
\begin{tabular}{l l c c c c c c} \hline
$~~~-(T+C)$ & $~~~-(C-P)$         & $-(T+P)$     &
  $(P)$               & \\ \hline
$\s(B^+\to\pi^+\pi^0)$  & $\s(B^0\to\pi^0\pi^0)$ &
$B^0\to\pi^+\pi^-$ &
$B^+\to K^+ \Kbar^0$ & \\
                    & $\s(B_s\to\pi^0 \Kbar^0)$ & $B_s\to\pi^+K^-$   &
$B^0\to K^0 \Kbar^0$  &  \\ \hline
$~-(T'+C'+P')$ & $~~~-(C'-P')$ & $-(T'+P')$     &
  $(P')$               & \\ \hline
$\s(B^+\to\pi^0 K^+)$  & $\s(B^0\to\pi^0 K^0)$ & $B^0\to\pi^-
K^+$ &
$B^+\to\pi^+ K^0$ & \\
                       &                   & $B_s\to K^- K^+$  &
$B_s\to K^0 \Kbar^0$  &  \\ \hline
\end{tabular}
\end{center}
\end{table}

The three independent triangle relations and one quadrangle relation
are
\beq
\label{xyz}
(T+C)=(C-P)+(T+P)~,
\eeq
\beq
\label{xypzp}
(T+C)=(C'-P')/r_u + (T'+P')/r_u~,
\eeq
\beq
\label{xwp}
(T+C) = (T'+C'+P')/r_u - (P')/r_u~,
\eeq
\beq
\label{zp}
(T'+P') - (P') = r_u (T+P) - r_u(P)~.
\eeq
For example, by using Table 1 we can rewrite the relation in
Eq.~(\ref{xyz}) in terms of decay amplitudes as:
\beq
\label{trita}
\s A(B^+ \to \pi^+ \pi^0) = \s A(B^0 \to \pi^0 \pi^0) +
A(B^0 \to \pi^+ \pi^-) ~.
\eeq
We have chosen to express this relation using $B^0$ and $B^+$
mesons only.   However, by using the amplitude equalities from
Table 1, we could equally have written the right side of the above
relation in terms of $B_s$.

The surprising result \cite{HLGR} is that the three triangle relations
allow us to {\it completely} solve for the magnitudes and phases of the
amplitudes $T,C,P$. In addition we will have enough independent
determinations of the same quantities to be able to test our two
assumptions, namely SU(3) symmetry and the neglect of the $E,A,PA$
diagrams.

Since the amplitude for $B\to\pi^+\pi^0$ decay, given by $-(T+C)/\s$, is
pure $\Delta I = 3/2$, the diagram $(T+C)$ has only one term, which we
denote by $A_{\sss I=2} e^{i\phi_2} e^{i\delta_2}$. Thus, for example, the
triangle relation given in Eq.~(\ref{xyz}) becomes
\beq
A_{\sss I=2} e^{i\phi_2} e^{i\delta_2}
=(A_C e^{i\phi_{C}}e^{i\delta_C}-A_P e^{i\phi_P}e^{i\delta_P})
+(A_T e^{i\phi_T}e^{i\delta_T}+A_P e^{i\phi_P}e^{i\delta_P})~~~,
\eeq
and similarly for the other relations. As before, the $\phi_i$ are the weak
phases and the $\delta_i$ are the strong phases. The $\delta_i$ are chosen
such that the quantities $A_{\sss I=2}$, $A_T$, $A_{T'}$, $A_C$, $A_{C'}$,
$A_P$ and $A_{P'}$ are real and positive (only relative strong phase
differences are physically meaningful). SU(3) symmetry implies that the
strong phases for the primed and unprimed graphs are equivalent. Working
within the Wolfenstein approximation of the CKM matrix, it is easy to see
that the weak phases of the various amplitudes are: $\phi_2= \phi_T =
\phi_{T'} = \phi_C = \phi_{C'} = \gamma$, $\phi_P=-\beta$, and $\phi_{P'} =
\pi$ (up to corrections of order $\lambda^2 \approx 0.05$). Also,
$A_{T'}/r_u=A_T$ and $A_{C'}/r_u=A_C$. Finally, multiplying through on both
sides by $\exp(-i\gamma-i\delta_2)$, the 3 triangle relations become
\begin{eqnarray}
\label{taa}
A_{\sss I=2} & = &
(A_C e^{i\Delta_C} + A_P e^{i\alpha} e^{i\Delta_P})
+(A_T e^{i\Delta_T} - A_P e^{i\alpha} e^{i\Delta_P}), \\
\label{tb}
A_{\sss I=2} & = &
(A_C e^{i\Delta_C} + A_{P'} e^{-i\gamma}e^{i\Delta_P}/r_u)
+(A_T e^{i\Delta_T} - A_{P'} e^{-i\gamma}e^{i\Delta_P}/r_u), \\
\label{tc}
A_{\sss I=2} & = & (A_T e^{i\Delta_T} + A_C e^{i\Delta_C}
- A_{P'} e^{-i\gamma}e^{i\Delta_P}/r_u)
+ A_{P'} e^{-i\gamma}e^{i\Delta_P}/r_u),
\end{eqnarray}
where we have defined $\Delta_i \equiv \delta_i - \delta_2$.

Consider first the two triangle relations in Eqs.~(\ref{tb}) and
(\ref{tc}). These relations define two triangles which share a common base.
Each triangle is determined up to a two-fold ambiguity, since it can be
reflected about its base. Implicit in these two triangle relations is the
relation
\beq
\label{cttri}
A_{\sss I=2} = |T + C| = A_T e^{i\Delta_T} + A_C e^{i\Delta_C}~~~.
\eeq

\begin{figure}
\centerline{\psfig{figure=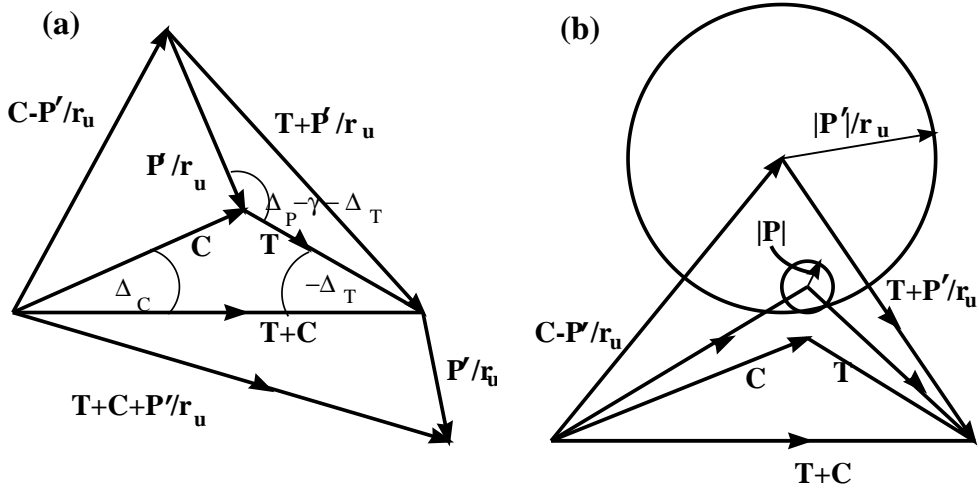,height=7.0cm,angle=0}}
\caption{\tenrm\baselineskip=12pt Triangle relations used to obtain weak
phases and strong final-state phase shift differences. (a) Relation based
on Eqs.~(\protect\ref{cttri}) (upper triangle) and (\protect\ref{tc})
(lower triangle). (b) Relation based on Eqs.~(\protect\ref{taa}) (lower
triangle with small circle about its vertex) and (\protect\ref{tb}) (upper
triangle with large circle about its vertex). The relation based on
(\protect\ref{taa}) and (\protect\ref{tc}) follows an almost identical
construction.}
\label{twotri}
\end{figure}

Thus both of these triangles also share a common subtriangle with sides
$T+C$, $C$ and $T$ as shown in Fig.~\ref{twotri}. The key point is this: the
subtriangle is completely determined, up to a four-fold ambiguity, by the
two triangles in Eqs.~(\ref{tb}) and (\ref{tc}). This is because both the
magnitude and relative direction of $P'/r_u$ are completely determined by
constructing the triangle in Eq.~(\ref{tc}). Therefore the point where the
vectors $C$ and $T$ meet is given by drawing the  vector $P'/r_u$ from the
vertex opposite the base [see Fig.~\ref{twotri}]. (A similar construction
would have given the same point if we had used the vector $T+P'/r_u$
instead of $P'/r_u$.) Thus by measuring the five rates for

\vspace*{2mm}
$B^0 \to \pi^0 K^0$ (giving $|C - P'/r_u|$),

\vspace*{2mm}
$B^0 \to \pi^-K^+$ (giving $|T + P'/r_u|$),

\vspace*{2mm}
$B^+ \to \pi^0 K^+$ (giving $|T + C + P'/r_u|$),

\vspace*{2mm}
$B^+ \to \pi^+ K^0$ (giving $|P'/r_u|$), and

\vspace*{2mm}
$B^+ \to \pi^+\pi^0$ (giving $|T + C|=A_{\sss I=2}$,
i.e.~the triangle's base),

\vspace*{2mm}
\noindent
we can determine $\Delta_P - \gamma$, $|T|$ and $|C|$, up to a two-fold
ambiguity and $\Delta_C$ and $\Delta_T$ up to a four-fold ambiguity. As we
will discuss later, these discrete ambiguities can be at least partially
removed through the knowledge of the relative magnitudes of $|P|$, $|C|$,
$|T|$ and $|P'|$, and through independent measurements of the amplitudes
and the strong and weak phases.

If we also measure the rates for the CP-conjugate processes of the above
decays, we can get more information. These CP-conjugate decays obey similar
triangle relations to those in Eqs.~(\ref{tb}) and (\ref{tc}). However,
recall that under CP conjugation, the weak phases change sign, but strong
phases do not. Thus we can perform an identical analysis with the
CP-conjugate processes, giving us another, independent determination of
$|T|$, $|C|$, $\Delta_C$ and $\Delta_T$. But, instead of $\Delta_P -
\gamma$, this time we get $\Delta_P + \gamma$. Thus we obtain $\Delta_P$
and $\gamma$ separately. Note that it is not, in fact, necessary to measure
all 5 CP-conjugate processes. The rate for $B^- \to \pi^-\pi^0$ is the same
as that for $B^+ \to \pi^+\pi^0$, since they involve a single weak phase
and a single strong phase. Similarly, the rates for $B^+ \to \pi^+ K^0$ and
$B^- \to \pi^- K^0$ are equal. Therefore, in order to extract $\gamma$, in
addition to the above 5 rates, we need only measure
%
%\vspace*{2mm}
$\Bbar^0 \to \pi^0 \Kbar^0$ (giving $|\bar C - \bar{P'}/r_u|$),
%
%\vspace*{2mm}
$\Bbar^0 \to \pi^+K^-$ (giving $|\bar T + \bar{P'}/r_u|$), and
%
%\vspace*{2mm}
$B^- \to \pi^0 K^-$ (giving $|\bar T + \bar C + \bar{P'}/r_u|$).
%
%\vspace*{2mm}
%\noindent
To sum up, by measuring the above 8 rates, the following quantities can be
obtained: the weak phase $\gamma$, the strong phase differences $\Delta_T$,
$\Delta_C$ and $\Delta_P$, and the magnitudes of the different amplitudes
$|T|,~|C|$ and $|P'|$.

Note that the two triangles given by the relations in Eqs.~(\ref{taa}) and
(\ref{tb}) share a common base with each other and also with the
sub-triangle in Eq.~(\ref{cttri}) (which still holds). The same is true for
the two triangles constructed using  the triangle relations in Eqs.\
(\ref{taa}) and (\ref{tc}). Unlike the first two-triangle construction,
however, the shape of the sub-triangle is not yet fixed. Nevertheless, the
point where the vectors $C$ and $T$ meet can still be determined by
measuring the additional decays represented by $P$, $P'$, or $|T+P'/r_u|$.
A detailed explanation of these two constructions can be found in
Ref.~\cite{HLGR}. The point is that by measuring 7 rates  we can extract
$\Delta_P+\alpha$, $\Delta_P - \gamma$, $\Delta_C,$ and $\Delta_T$, up to
an eight-fold ambiguity, and $|T|$ and $|C|$ up to a four-fold ambiguity.
Through the two quantities  $\Delta_P+\alpha$ and $\Delta_P - \gamma$, we
can then determine the weak phase $\beta$ (using
$\beta=\pi-\alpha-\gamma$), up to discrete ambiguities. As in the first
two-triangle construction, all rates are time-independent. What is
surprising, perhaps, about this particular construction is that {\it it is
not even necessary to measure the CP-conjugate rates in order to obtain
$\beta$.} The reason is that SU(3) flavor symmetry implies the equality of
the strong final-state phases of two different amplitudes, in this case $P$
and $P'$. Subtracting the (strong plus weak) phase of one amplitude from
the other then determines a weak phase. Usually, in a given process,
without measuring the charge-conjugate rate one can only measure the sum of
a weak and a strong phase.

If the CP-conjugate rates are also measured, we can obtain $\Delta_P$,
$\alpha$, and $\gamma$ separately. This provides another, independent
determination of $|T|$, $|C|$, $\Delta_C$ and $\Delta_T$. As in the first
construction, no observation of CP violation is necessary to make such
measurements. Again, it is not necessary to measure all the CP-conjugate
rates -- only four can be different from their counterparts.

\section{Testing Our Assumptions}
The three constructions use $B$ decays to $\pi\pi$, $\pi K$ and $K{\bar K}$
final states. At present, the decays $B^0 \to \pi^+\pi^-$ and/or $\pi^-
K^+$ have been observed, but the two final states cannot be distinguished
\cite{CLEO}. The combined branching ratio is about $2\times 10^{-5}$.
Assuming equal rates for $\pi^+\pi^-$ and $\pi^- K^+$, which seems likely,
the amplitudes $|T|$ and $|P'|$ should be about the same size. On the other
hand, the amplitude $|C|$ is expected to be about a factor of 5 smaller:
the amplitudes $|T|$ and $|C|$ are basically the same as $|a_1|$ and
$|a_2|$, respectively, introduced in Ref.~\cite{BSW}, for which the values
$|a_1| = 1.11$ and $|a_2| = 0.21$ have been found \cite{Lindner}. The ratio
$|P/T|$ has also been estimated to be small, $\lsim 0.20$
\cite{gropenguin}. Therefore all the decays used in these constructions
should have branching ratios of the order of $10^{-5}$, with the exception
of $B\to K{\bar K}$ ($P$) and $B^0\to\pi^0\pi^0$ [$\sim (C-P)$], which are
probably an order of magnitude smaller.

The knowledge that the amplitudes obey the hierarchy
$|P|,~|C|<|T|<|P'/r_u|$ will also help in reducing discrete ambiguities.
For example, in the first two-triangle construction [Fig.~\ref{twotri}], we
noted in the discussion following Eq.~(\ref{cttri}) that the subtriangle
can be determined up to a four-fold ambiguity. However, two of these four
solutions imply that $|C|$ and $|T|$ are both of order $|P'/r_u|$, which
violates the above hierarchy. Thus the four-fold ambiguity in the
determination of the subtriangle is reduced to a two-fold ambiguity, and
the discrete ambiguities in the determination of subsequent quantities such
as $\Delta_P-\gamma$, $\Delta_C$, etc., are likewise reduced. The
ambiguities in the other two constructions can be partially removed in a
similar way.

All three two-triangle constructions described above rely on two
assumptions. The first is that the diagrams $A$, $E$ and $PA$ (and their
primed counterparts) can be neglected. This can be tested experimentally.
The decays $B^0 \to K^+ K^-$ and $B_s \to \pi^+ \pi^-$ can occur only
through the diagrams $E$ and $PA$, and $E'$ and $PA'$, respectively.
Therefore, if the above assumption is correct, the rates for these two
decays should be much smaller than the rates for the decays in Table 1.

The second assumption is that of an unbroken SU(3) symmetry. We know,
however, that SU(3) is in fact broken in nature. Assuming factorization,
SU(3)-breaking effects can be taken into account by including the meson
decay constants $f_\pi$ and $f_K$ in the relations between $B\to\pi\pi$
decays and $B\to\pi K$ decays \cite{Silvawolf}. In other words, the factor
$r_u$ which appears in two of the triangle relations should be multiplied
by $f_K/f_\pi\approx 1.2$. One way to test whether this properly accounts
for all SU(3)-breaking effects is through the rate equalities in Table 1.
Even if it turns out that $f_K/f_\pi$ does not take into account all
SU(3)-breaking effects, the large number of independent measurements is
likely to help in reducing uncertainties due to SU(3) breaking. For
example, note that, not counting the CP-conjugate processes, the last two
constructions have six of their seven rates in common. This means that a
measurement of only eight decay rates gives two independent measurements of
$|T|$, $|C|$, $\Delta_C$, $\Delta_T$, $\Delta_P - \gamma$ and $\Delta_P +
\alpha$. In fact, these eight rates already contain the five rates of the
first construction [Fig.~\ref{twotri}]. Thus we actually have three
independent ways of arriving at $|T|$, $|C|$, $\Delta_C$, $\Delta_T$ and
$\Delta_P - \gamma$. Including also the CP-conjugate processes, we have a
total of 13 $B$-decay rate measurements which give us six independent ways
to measure $|T|$, $|C|$, $\Delta_C$ and $\Delta_T$, five ways to measure
$\Delta_P$, three independent ways to measure $\gamma$, and two ways to
measure $\alpha$. (If time-dependent measurements are possible, there are
additional independent ways to measure $\alpha$.) The point is that the
three two-triangle constructions include many ways to measure the same
quantity. This redundancy provides a powerful way to test the validity of
our SU(3) analysis and reduces the discrete ambiguities in the
determination of the various quantities.

A simpler system where a subset of
these assumptions can be tested are the decays of $B$'s to one light
pseudocalar and one charmed meson\cite{GHLRii}.  Here one can also test for the
absence of exchange and annihilation graphs; there is no analogue of
the penguin annihilation graph.  Furthermore, the effects of decay
constants and form factors in SU(3) breaking can be studied
individually, whereas they occur together when both final-state mesons
are light.

Assuming that exchange and annihilation contributions can be neglected,
the following decay rates are expected to be equal:
\bigskip

\noindent (I)
$V_{cb}^* V_{ud} \sim {\cal O}(\lambda^2)$ processes:

(a) $B^0\to \pi^+ D^- = T+E$ and $B_s\to \pi^+ D^-_s = T$;

(b) $\sqrt{2}(B^0 \to \pi^0 \bar D^0) = C - E$ and $B_s \to \bar K^0 \bar D^0 =
C$;
\bigskip

\noindent (II)
$V_{cb}^* V_{us} \sim {\cal O}(\lambda^3)$ processes:

$B^0 \to K^+ D^- = T'$ and $B_s \to K^+ D_s^- = T' + E'$;
\bigskip

\noindent (III)
$V_{ub}^* V_{cs} \sim {\cal O}(\lambda^3)$ processes:

(a) $B^+ \to K^+ D^0 = -(\tilde C + \tilde A)$
and $B^0 \to K^0 D^0 = - \tilde C$;

(b) $B^0 \to \pi^- D_s^+ = - \tilde T$,
$\sqrt{2}(B^+ \to \pi^0 D_s^+) = - \tilde T$,
and $B_s \to K^- D_s^+ = - (\tilde T + \tilde E)$;
\bigskip

\noindent (IV)
$V_{ub}^* V_{cd} \sim {\cal O}(\lambda^4)$ processes:

(a) $\sqrt{2} (B^+ \to \pi^0 D^+)= - \tilde T' + \tilde A',~
B^0 \to \pi^- D^+ = -(\tilde T' + \tilde E')$,
and $B_s \to K^- D^+ = - \tilde T'$;

(b) $B^+ \to \pi^+ D^0 = -(\tilde C' + \tilde A'),~
\sqrt{2}(B^0 \to \pi^0 D^0)= - \tilde C' + \tilde E'$,
and $B_s \to \bar K^0 D^0 = - \tilde C'$.
\bigskip

Here in the SU(3) limit $T'/T = C'/C = E'/E
= \tilde T'/ \tilde T = \tilde C'/ \tilde C = \tilde E' / \tilde E = \tilde A'
/ \tilde A = |V_{us}/V_{ud}| = |V_{cd}/V_{cs}| = r_u=0.23$.

The following processes are expected to be suppressed by a term of order
$(f_B/m_b)$ with respect to the color-favored processes of the same order in
$\lambda$:
\bigskip

\noindent (I)
$V_{cb}^* V_{ud} \sim {\cal O}(\lambda^2)$ processes:

$B^0\to K^+ D^-_s = E$.
\bigskip

\noindent (II)
$V_{cb}^* V_{us} \sim {\cal O}(\lambda^3)$ processes:

$B_s \to \pi^+ D^- = E'$ and $-\sqrt{2}(B_s \to \pi^0 \bar D^0) = E'$.
\bigskip

\noindent (III)
$V_{ub}^* V_{cs} \sim {\cal O}(\lambda^3)$ processes:

$B^+\to K^0 D^+ = \tilde A,~-(B_s \to \pi^- D^+) = \tilde E$, and
$\sqrt{2}(B_s \to \pi^0 D^0) = \tilde E$.
\bigskip

\noindent (IV)
$V_{ub}^* V_{cd} \sim {\cal O}(\lambda^4)$ processes:

$B^+\to \bar K^0 D^+_s = \tilde A'$ and $B^0 \to K^- D_s^+ = - \tilde E'$.
\bigskip

The effect of form factors in SU(3) symmetry breaking can be directly
studied by comparing spectator quark processes in which strange and
non-strange quarks combine with strange or non-strange quarks.
For example consider the following ratios of rates.
\beqn
\Gamma(B^o \to K^+ D^-)/\Gamma(B_s \to \pi^+ D_s^-) &=& |T'/T|^2 \\
\label{one}
& &  \nonumber \\
\Gamma(B^o \to K^0 \bar D^0)/\Gamma(B_s \to \bar K^0 \bar D^0) &=& |C'/C|^2\\
  \label{two}
& & \nonumber \\
\Gamma(B_s \to K^- D^+)/\Gamma(B^0 \to \pi^- D_s^+)&=&|\tilde T'/\tilde T|^2\\
   \label{three}
& &  \nonumber \\
\Gamma(B_s \to \bar K^0 D^0)/\Gamma(B^0 \to K^0 D^0)&=&|\tilde C'/\tilde C|^2
 \label{four}
\eeqn

In the absence of SU(3) symmetry breaking  they should all equal
$r_u^2=|V_{us}/V_{ud}|^2$.  (We are not interested in isospin
symmetry breaking effects such as the deviation of the ratio
$\Gamma(B^+ \to \pi^0 D^+_s)/\Gamma(B^0\to \pi^- D^+_s)$ from 1/2)
Deviations in $r_u^2$
in    Eq.~(\ref{one}) will tell us about form factor effects in strange versus
non-strange combining with a heavy SU(3) singlet quark, whereas
Eq.~(\ref{three})
is the same spectator process combing with a non-strange SU(3)
anti-triplet.   Deviations in   Eq.~(\ref{two}) and
Eq.~(\ref{four}) will measure the same thing;
strange and non-strange combining with a non-strange and strange
repectively. Thus if form  factors are the main SU(3) breaking effects in
Eq.~(\ref{two}) and Eq.~(\ref{four}), we would expect equal but opposite
deviations from
$\lambda^2$ in these two processes.

Finally deviations in the following triangle relations test form
factor SU(3) symmetry breaking effects:
\beqn
 {\cal O}(\lambda^2)  {\rm process:\ \ }
(B^+ \to \pi^+ \bar D^0)  &=&   (B_s \to \pi^+ D_s^-)  + (B_s \to \bar K^0 \bar
D^0) \\  \label{five}
 (T+C)        &=&   ~~~~~~~~~~~ ( T ) ~~  +  ~~ ( C )   \nonumber \\
 {\cal O}(\lambda^3) {\rm process:\ \  }
(B^+ \to K^+ \bar D^0)  &=& (B^0 \to \bar K^+  D^-) + (B^0 \to K^0  \bar D^0)
\\  \label{six}
 (T+C)        &=&   ~~~~~~~~~~~ ( T'  ) ~~  +  ~~ ( C' )     \nonumber
\eeqn
where, for example,  on the left-hand side of
Eq.(\ref{five}) we have non-strange quarks
combining with non-strange quarks, whereas on
the right-hand side we have strange combining
with non-strange.

\section{Summary and Conclusions}

We have also described in some detail the recent developments which
provide a prescription for the measurement of all relevant quantities:
weak and strong phases, and the sizes of the contributing diagrams.
This analysis uses SU(3) flavour symmetry along with the important
dynamical assumption that exchange and annihilation diagrams can be
neglected. This method relies on several triangle relations which hold
under these assumptions.  Like $B\to DK$ decays, neither
time-dependent measurements nor tagging are required. This analysis
can therefore be carried out at a symmetric $B$-factory such as CLEO.
Unlike $B\to DK$, however, the branching ratios for most of the
processes involved are expected to be $O(10^{-5})$, so that the sides
of the triangles are all roughly the same size. This method also
provides enough redundancy to test the consistency of the assumptions.

\section{Acknowledgements}  I wish to thank the people I learned
B-physics from especially my  collaborators Michael Gronau, Brian
Hill, David London, and Jonathan Rosner.  I also wish to express my
gratitued to the organizers  especially Miguel Angel Perez and Matias
Moreno.

\end{document}